\documentclass{cfm2011}
\usepackage[latin1]{inputenc}
\usepackage{graphicx}
\usepackage{color}

\title{Modélisation et évaluation du comportement dynamique d'une structure de machine outil UGV pour la prédiction des défauts géométriques}
\author{D. Prévost\inst{a}, S. Lavernhe\inst{a}, F. Louf\inst{b}, P. Rouch\inst{c}}

\instlabel{a}{LURPA, ENS Cachan, 61 avenue du Président Wilson, 94235 CACHAN}
\instlabel{b}{LMT, ENS Cachan, 61 avenue du Président Wilson, 94235 CACHAN}
\instlabel{c}{LBM, Arts et Métiers Paristech, 151 Boulevard de l'hôpital, 75013 PARIS}

\begin{document}

\maketitle

\begin{resume}
	 \textit{Pour atteindre des performances accrues en terme de productivité, les centres d'Usinage Grande Vitesse possèdent des caractéristiques cinématiques élevées. Les fortes accélérations mises en jeu lors de l'exécution des trajectoires sollicitent dynamiquement la structure et les liaisons. Les déformations engendrées peuvent alors se répercuter sur la position effective de l'outil par rapport à la pièce créant des écarts géométriques par rapport à la géométrie attendue.\\
Dans un contexte de maîtrise du processus de réalisation de pièces en UGV, les travaux développés présentent une démarche et un choix de modélisation éléments finis pour évaluer le comportement dynamique d'une structure de centre UGV. Basé sur ce modèle, une méthodologie de calcul est proposée pour évaluer les écarts géométriques générés lors de l'exécution de trajectoires types. Ces travaux constituent ainsi une voie pour le dimensionnement de structure, le réglage et l'optimisation des paramètres cinématiques pour l'usinage.}
\end{resume}

\begin{abstract}
	\textit{To achieve better performances in term of productivity, High Speed Machining centers have increasing kinematical characteristics. High accelerations caused by the trajectory execution dynamically load mechanical structure, axes and joints. Such deformations of mechanical components affect the actual position of the tool in the machining frame. Hence, these gemetrical deviations make the machined geometry of the part vary from the expected one.\\ 
Within the context of managing the HSM process, this study develops a framework and choice of finite element model to evaluate the structure dynamical behavior of a HSM center. Based on the proposed model, a calculation method is detailed to evaluate the geometrical deviations induced by the trajectory execution. The presented works can thus be used as a tool for the structure design of machine, the adjustment and the optimization of kinematical parameters for machining.}
\end{abstract}

\keywords{Usinage Grande Vitesse ; dynamique de structure ; écarts géométriques}


\section{Introduction}

L'Usinage à Grande Vitesse est souvent utilisé pour réaliser des pièces de formes gauches à forte valeur ajoutée telles que les moules, matrices, aubes de turbines où la géométrie complexe est un des paramètres de haute technicité à obtenir. L'obtention de telles surfaces usinées passe par la maîtrise du processus UGV où il est nécessaire de maîtriser chacune des étapes, tant d'un point de vu local que sur le déroulement global du processus pour contrôler le résultat final. Classiquement le processus est découpé en trois activités (fig.~\ref{image_processus}) :
\begin{figure}[htp]
	\center{\scalebox{0.5}{\includegraphics* {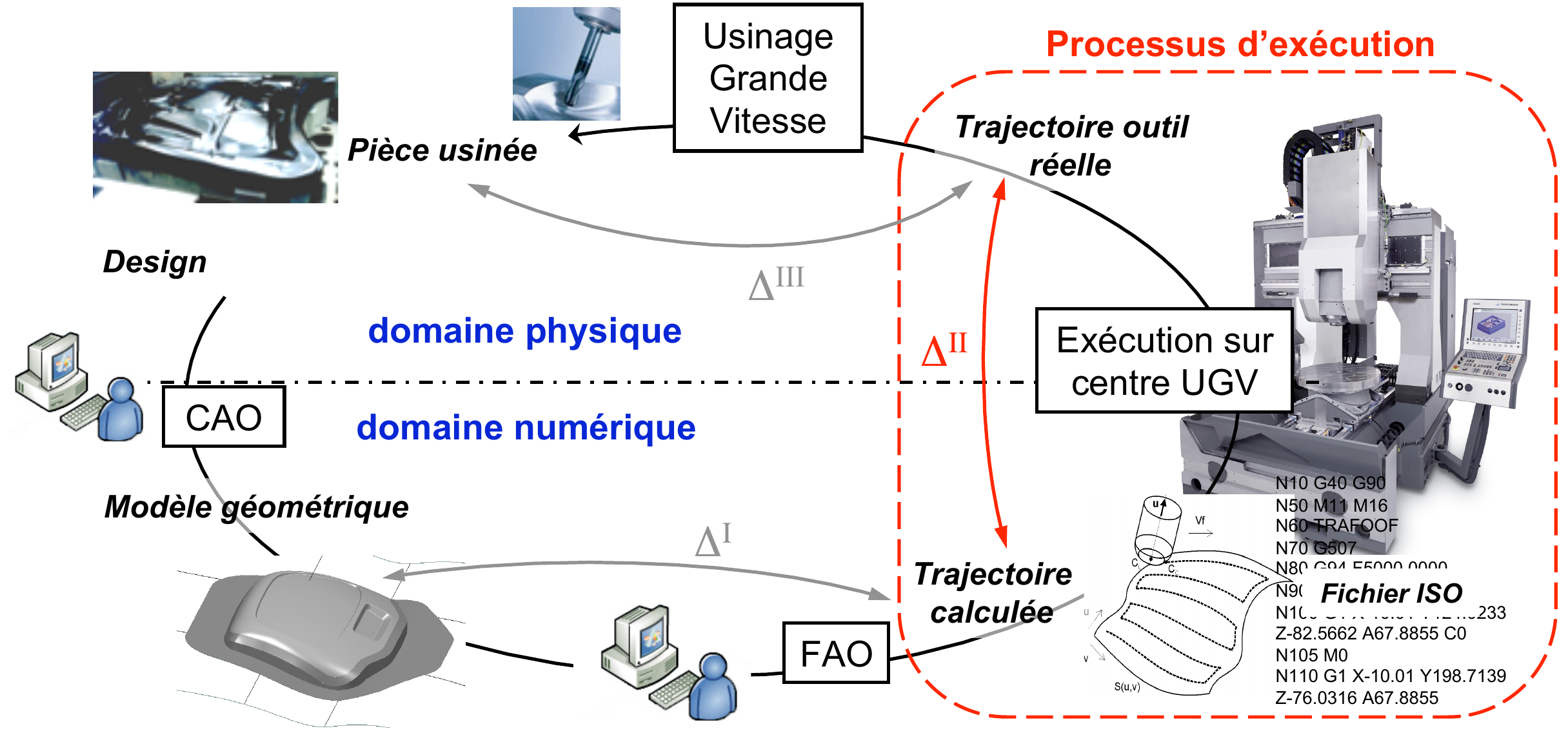}}}
          \caption{Processus d'usinage grande vitesse et écarts induits}
          \label{image_processus}
\end{figure}\\
La première étape dite de génération de trajectoire consiste à calculer un ensemble de positionnements outils sur le modèle géométrique de la pièce considéré comme référence. La trajectoire outil/pièce est alors décrite dans un fichier par l'interpolation de ces positionnements selon le format de description choisi (linéaire ou polynomial) interprétable par la Commande Numérique (CN) de centre UGV.\\
L'exécution des trajectoires définies en FAO est ensuite réalisée par le couple CN, machine-outil. Le rôle de la CN est, à partir du programme d'usinage, de générer en temps réel les consignes de position de chaque axe machine. La trajectoire est à ce stade re-discrétisée géométriquement puis suivie temporellement selon des profils et caractéristiques cinématiques prédéfinis pour obtenir un ensemble de mouvements d'axes régulés par les moteurs et variateurs au sein d'un asservissement. Le mouvement outil-pièce résultant est enfin donné par la structure physique de la machine (disposition relative des axes, les mouvements de axes ou liaisons, etc...).
La troisième étape du processus est relative à la coupe, interaction outil-pièce pour générer la géométrie usinée.\\
Chacune de ces trois étapes induit des écarts géométriques (notés respectivement $\Delta^{I}$,$\Delta^{II}$ et $\Delta^{III}$) entre la surface générée (ou l'enveloppe de la trajectoire associée) et la surface de référence pour l'étape considérée. Pour maîtriser la géométrie finale, il est donc nécessaire de contrôler les diverses sources de défauts. De nombreux développements sont aujourd'hui menés localement pour optimiser la trajectoire outil, l'interpolation, le suivi ou encore la géométrie machine, etc... aboutissant alors à des solutions partielles ne garantissant pas le bon déroulement du processus global. Contenir le défaut géométrique final et maximiser la productivité nécessite une intégration des différents composants et transformations au sein de toutes les étapes, depuis la définition géométrique de la pièce à la coupe.

\begin{figure}[htp]
	\center{\scalebox{0.7}{\includegraphics* {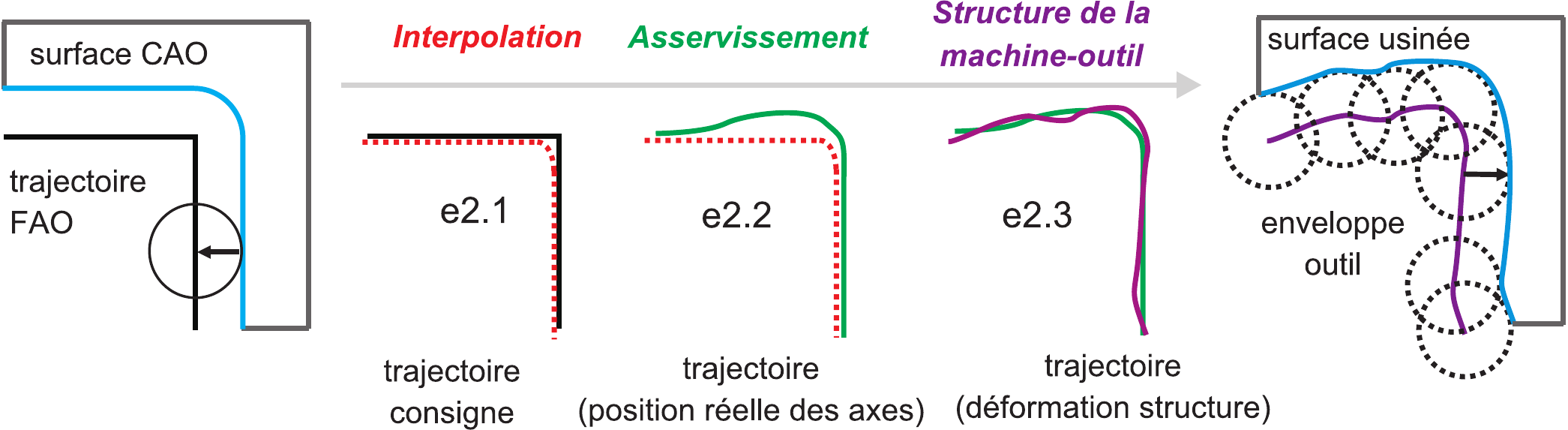}}}
          \caption{Détail des écarts induits sur la trajectoire au cours du processus d'exécution}
          \label{image_ecarts}
\end{figure}

Dans ce contexte, les écarts géométriques induits sur la trajectoire (fig. \ref{image_ecarts}) par l'interpolation temps réel (écart e2.1) et ceux liés au suivi de chaque axe (écart e2.2) ont été préalablement investigués dans \cite{Lavernhe08} et \cite{Prevost10}. Il est ainsi possible de réduire la somme vectorielle des écarts liés à l'interpolation et à leur suivi ($\delta 2.1+\delta 2.2$) en anticipant les défauts par le calcul d'une trajectoire déformée.
\\
Les travaux développés portent sur le comportement dynamique des machines-outils et de leur impact sur la géométrie usinée. Ils se focalisent sur le lien entre les trajectoires d'usinage exécutées à grande vitesse et les écarts géométriques engendrés par la structure de la machine (écart e2.3 fig. \ref{image_ecarts}). En effet, dans le cadre de la finition, les surépaisseurs enlevées restant très faibles sur les formes gauches, les actions mécaniques de coupe et déformations associées peuvent être négligées ; cependant les longueurs de trajectoires restant très importantes, il est nécessaire de les suivre avec une vitesse très élevée pour rester productif.

La section suivante est consacrée à la démarche de choix et de modélisation mise en place pour les structures de centre UGV dites sérielles, sur lequelles portent l'étude. Le modèle est ensuite exploité pour venir évaluer le comportement dynamique de la structure et le déplacement résultant de la pointe outil lors d'une excitation donnée par le suivi de trajectoire.


\section{Modélisation du centre UGV}

\subsection{Modélisation géométrique et mécanique}

\begin{figure}[htp]
	\center{\scalebox{0.5}{\includegraphics* {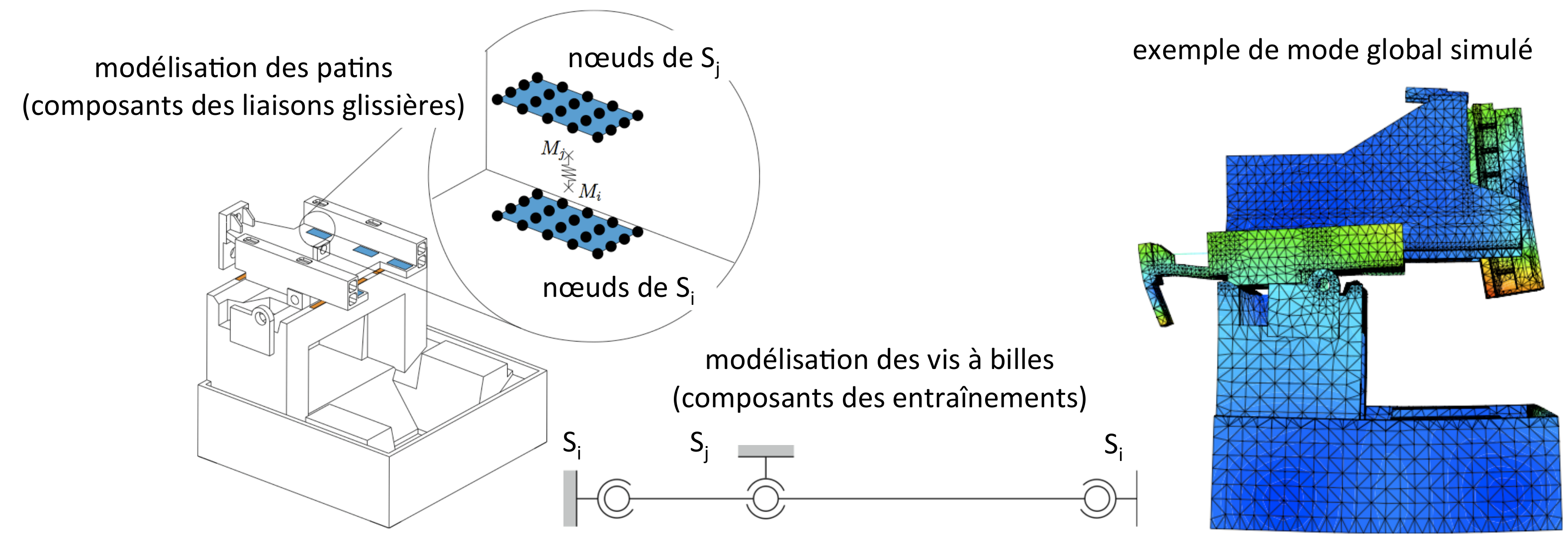}}}
          \caption{Eléments de modélisation et exemple de mode simulé}
          \label{image_modele}
\end{figure}

L'étude porte sur le centre UGV 5 axes Mikron UCP710 du LURPA de type RRTTT en considérant dans un premier temps uniquement les trois mouvements de translation associés aux axes X, Y et Z. Les différentes techniques de modélisation des structures pour la simulation des écarts géométriques sous chargement dynamique peuvent être regroupées en quatre catégories, avec chacune ses avantages et inconvénients : de type masses concentrées, de type multi-corps solides rigides reliés par des liaisons souples, de type Eléments Finis (E.F) ou de type hybride mélangeant les trois précédentes \cite{Altintas05}\cite{Weule03}\cite{Zaeh07}. Sur la structure étudiée, la complexité géométrique des solides et la souplesse induite par les nombreuses liaisons nécessite une approche hybride associant la déformation des corps par E.F. et la modélisation de toutes les liaisons. Ainsi, le modèle est constitué de (fig. \ref{image_modele}):
\begin{itemize}
	\item une modélisation E.F. pour chaque sous ensemble de solide correspondant aux axes;
	\item des liaisons élastiques linéaires au niveau de chaque patin de liaison glissière;
	\item un modèle poutre pour chaque entraînement par vis à bille;
	\item des masses concentrées pour des composants et éléments spécifiques de la structure. 
\end{itemize}

La modélisation des solides est réalisée sur une géométrie mesurée et paramétrée pour représenter l'ensemble des configurations machine correspondant au volume de travail. Les volumes sont maillés à l'aide de tétraèdres. L'interpolation utilisée est de type linéaire.

Chaque liaison glissière de direction $\vec{n}_{ij}$ est réalisée par 6 patins à rouleaux préchargés disposés sur deux rails parallèles. Chacun de ces patins peut être modélisé localement comme une liaison sphère-cylindre élastique entre les axes. Du point de vue éléments finis, ces liaisons élastiques entre les degrés de libertés sont imposées avec une vision maître esclave.

Chaque vis à billes est modélisée comme une poutre et maillée à l'aide d'éléments linéaires. Les liaisons entre chaque vis à billes et les deux axes qu'elle relie sont considérées comme rigides. La liaison hélicoïdale entre l'écrou et la vis est modélisée comme une rotule de sorte qu'il n'y a pas de couplage entre la rotation de la vis et la translation de l'écrou.

La broche logée dans l'axe Z, la table (non représentée) et les moteurs actionnant les vis à billes sont modélisés par une rigidification de leur support et une masse. Les inerties de ces différents éléments ne sont pas prises en compte. La liaison de la machine avec le sol est réalisée à l'aide de patins isolants. Ces patins sont modélisés par une rigidification de la zone du bâti en contact avec eux et un appui élastique possédant trois raideurs en translation et aucune raideur en rotation.

\subsection{Analyse modale et validation du modèle}

Une fois le modèle défini, et pour un jeu de paramètres structuraux donnés, les modes de la structure complète sont calculés dans une bande de fréquences comprises entre 0 et 400 Hz. Les modes observés dans cette bande de fréquences peuvent être classés en 4 catégories :
\begin{itemize}
	\item les modes déformant uniquement les appuis élastiques;
	\item les modes globaux déformant la structure;
	\item les modes locaux déformant une partie d'un des axes (supports moteur notamment);
	\item les modes locaux déformant les vis à billes (modes doubles généralement). 
\end{itemize}

Une étude de convergence a été effectuée. Elle montre que les fréquences propres propres calculées varient très peu à partir d'une taille de maille de 20 mm. Le modèle final comprend alors environ 300000 degrés de liberté. 
La figure \ref{image_modele} donne l'exemple d'un mode calculé.  Dans la bande de fréquences [0-400Hz], 75 modes ont été calculés. Il est important de noter que tous ces modes ne pourront pas être excités lors de l'exécution de la trajectoire. En effet, pour chaque axe, le constructeur peut spécifier un ensemble de filtres sur les consignes de vitesse et de courant appliqués au moteur, et la structure globale de l'asservissement réalise un filtre de type passe-bas lissant le mouvement effectif de chaque axe.


\section{Comportement lors de l'exécution de trajectoire UGV}

\subsection{Modélisation de la sollicitation dynamique}

Afin d'évaluer les déviations centre outil engendrées par la déformation de la structure, les sollicitations choisies doivent représenter le comportement cinématique pour des trajectoires sollicitant fortement les axes, c'est à dire possédant des vitesses et accélérations élevées. Les trajectoires sont constituées de segments de grande longueur qui permettent pour un profil cinématique limité par le jerk (trapèze d'accélération) d'atteindre des niveaux d'accélération élevés localement.
En supposant que le comportement dynamique de la machine varie peu sur l'espace de travail balayé lors de chacune des phases d'accélération, chaque solide (Si) correspondant aux axes est soumis à un champ d'accélération donné orienté par la trajectoire outil/pièce et imposé par la motorisation des axes. On a donc :
\begin{equation}
	\vec{\Gamma}_{S_i/(0)} = \sum_{j=2}^{i}\ddot{u}_{d,j-1} \vec{n}_{j-1} \quad \forall i\in\{2,3,4\}
\end{equation}
En écrivant que le champ de déplacement recherché est la somme d'un ensemble de mouvements de corps rigides et d'un champ déformant la structure
\begin{equation}
	\vec{U}(M,t) = \vec{u}(M,t) +  \sum_{j=2}^{i} u_{d,j-1} \vec{n}_{j-1}
\end{equation}
on peut réécrire l'équation d'équilibre éléments finis sous la forme suivante :
\begin{equation}
	[K] \{u(t)\} + [C]\{\dot{u}(t)\} + [M]\{\ddot{u}(t)\} = -[M]\{\ddot{u}_d(t)\} 
	\label{equation_systdiff}
\end{equation}
où $\{\ddot{u}_d(t)\} $ est le champ d'accélération défini sur chacun des axes $(S_i)$ par les vecteurs $\vec{\Gamma}_{S_i/(0)}$. L'hypothèse est faite ici que les termes $[C]\{\dot{u}_d(t)\}$ sont négligeables devant les efforts d'inertie.\\

Afin de réduire le coût de calcul associé à la résolution du système différentiel (\ref{equation_systdiff}), le champ de déplacement $\{u(t)\}$ est cherché dans la base modale $[\Phi]$ de dimension $n_{ddl}\times n_{modes}$. On obtient de cette manière $n_{modes}$ équations différentielles découplées à résoudre \cite{Geradin96}. En introduisant une discrétisation temporelle ayant pour but de représenter la variation temporelle du second membre et la solution elle même, la résolution est explicite sur chaque pas de temps.

\subsection{Résultats}
\begin{figure}[htp]
	\center{\scalebox{0.4}{\includegraphics* {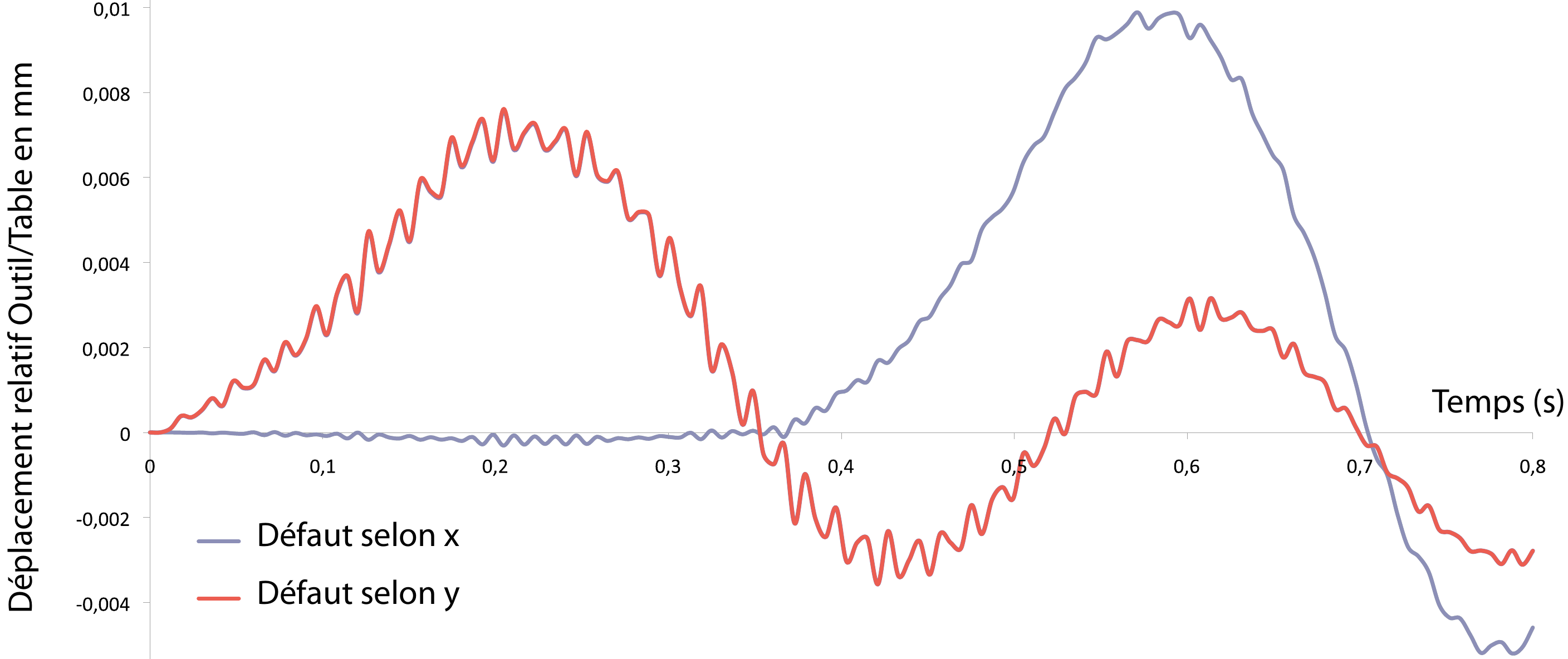}}}
          \caption{Projection des écarts centre-outil simulés pour un passage de coin}
          \label{image_ecartsXY}
\end{figure}

\begin{figure}[htp]
	\center{\scalebox{0.7}{\includegraphics* {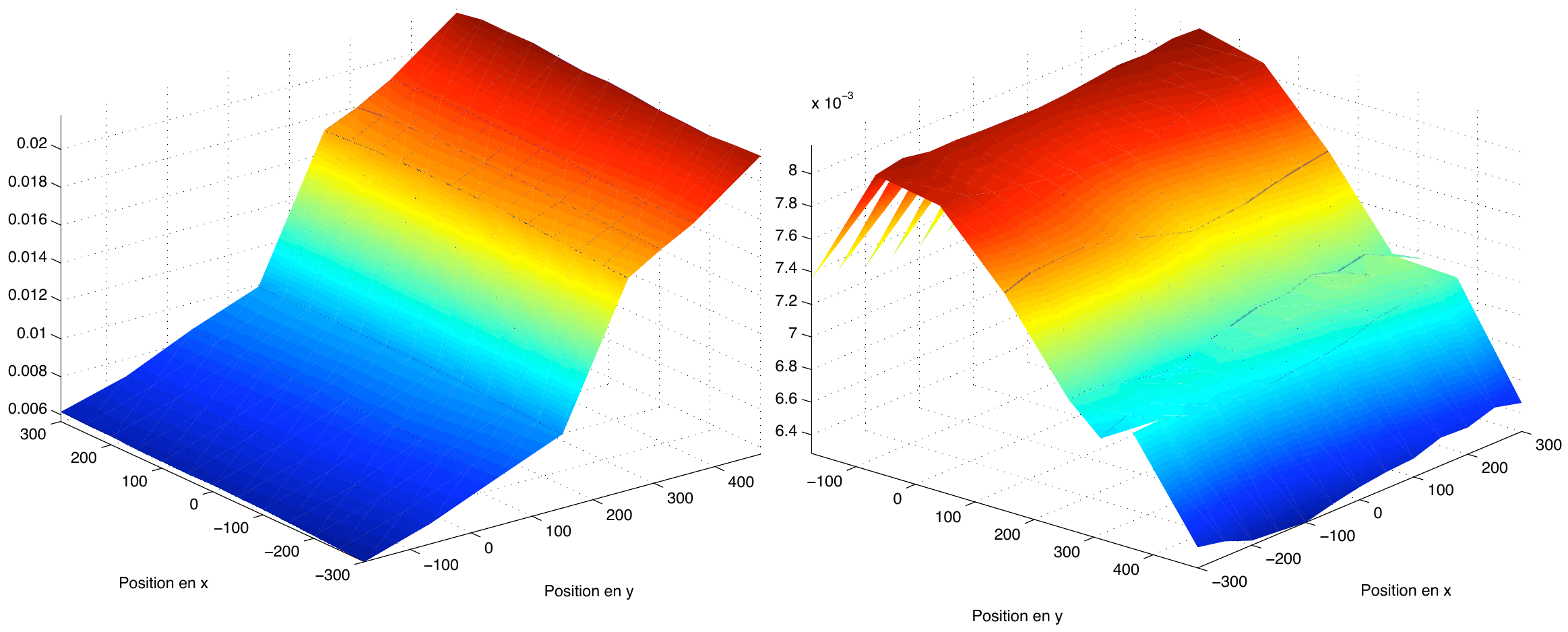}}}
          \caption{Composantes $\vec{x}$ (gauche) et $\vec{y}$ (droite) du défaut maximal en fonction de la position dans l'espace de travail (plan XY)}
          \label{image_ecartsmax}
\end{figure}

Ainsi, au voisinage d'une configuration articulaire donnée, le calcul donne les évolutions temporelles des déplacements du point centre de l'outil lié rigidement à la broche et d'un point géométriquement identique au précédent mais lié rigidement à la table. Le déplacement relatif de ces deux points donne alors l'image de l'écart outil/pièce au niveau de la trajectoire. Supperposés aux déplacements nominaux des axes, la figure \ref{image_ecartsXY} présente les écarts simulés pour le passage d'un coin de poche.
\\
Les oscillations de la courbe rouge correspondent à la décélération de l'axe Y à l'arrivée du coin, puis à t=0,35s., l'axe X démarre en accélérant (courbe bleue) alors que les oscillations de l'axe Y sont amorties progressivement.\\
Il est alors possible d'évaluer le défaut maximal engendré par la déformation de la structure en faisant varier la position de la broche dans l'espace de travail de la machine. La figure \ref{image_ecartsmax} présente l'évolution de la déviation maximale de la trajectoire réelle vis à vis de la trajectoire issue de l'interpolation par la CN. On remarque que le défaut calculé sur $\vec{x}$ ou sur $\vec{y}$ dépendent peu de la position de l'axe X. Par contre la position de l'axe Y apparaît fortement influente sur ces défauts. Ces résultats semblent cohérents avec la structure sérielle de la machine sur laquelle l'axe Y est en porte-à-faux.


\section{Conclusion et Perspectives}
L'objectif global des travaux concerne la minimisation des défauts géométriques engendrés par le calcul des trajectoires et leur exécution sur centre UGV. Les travaux développés dans cette communication se focalisent sur l'évaluation des écarts outil/pièce à la pointe de l'outil générés par la déformation de la structure sollicitée par les déplacements des axes lors du suivi des trajectoires. Une modélisation hybride des composants de la structure d'un centre UGV industriel mélangeant modèle poutre, E.F, liaisons élastiques et masses concentrées est présentée. Elle permet de mettre en évidence une certaine richesse modale dans la plage [0-400Hz] en simulant les modes globaux et mouvements des corps associés. Associée aux lois de commande utilisées pour l'exécution de trajectoire, la résolution du problème dynamique dans la base modale permet de simuler le déplacement du centre de l'outil au cours du temps. Il est alors possible de déterminer sur l'espace de travail de la machine le plus grand défaut de positionnement de l'outil par rapport à la pièce. Il faut néanmoins noter que le modèle utilisé mériterait d'être recalé à partir d'essais statiques et/ou dynamiques sur la structure réelle ou sur des composants. En effet, les paramètres de rigidité des patins, des liaisons glissières, ou des liaisons entre vis et axes sont relativement mal connus. Par ailleurs, les niveaux d'amortissement modaux sont eux aussi à déterminer expérimentalement. \\
Cette démarche donne ainsi un levier pour optimiser le processus global : avec une connaissance plus fine des paramètres liés aux matériaux (raideurs, amortissements, masses), il est maintenant possible de relier le comportement dynamique basses fréquences de la structure avec plusieurs traitements amonts. La géométrie de la trajectoire, les lois de commandes utilisées pour l'interpolation temps réel et les asservissements des axes sont autant de points d'entrée pour régler le comportement machine. D'autre part, un tel modèle peut également être utilisé au niveau de la conception des centres UGV pour dimensionner au plus juste la géométrie et la masse des axes afin de rigidifier la structure tout atteignant des caractéristiques cinématiques élevées gages de productivité.


\section*{Remerciements}
Les auteurs tiennent à remercier l'institut Farman de l'ENS Cachan pour le soutien financier de ces travaux pour le projet \textit{OPTRAJ} (http://www.farman.ens-cachan.fr/).\\
Ce travail est conduit sous l'égide du groupe de travail Manufacturing 21 qui comprend 18 laboratoires français. Les thèmes abordés sont la modélisation du processus de fabrication, l'optimisation des conditions opératoires, la fabrication virtuelle, la validation expérimentale et le développement de nouvelles méthodes de fabrication.


\end{document}